# Synchronized Opto-Electro-Mechanical Measurements for Energy Loss Estimation in Thin-Film-Piezoelectric-on-Substrate MEMS/NEMS Devices

Vishnu Kumar[1], Sudhanshu Tiwari[1,2], Gayathri Pillai[1], Rudra Pratap[1,3], Saurabh Chandorkar[1]

[1]Centre for Nano Science and Engineering, Indian Institute of Science, Bengaluru, India

[2]Elmore Family School of Electrical and Computer Engineering, Purdue University, USA

[3]Plaksha University, Mohali, Punjab, India

Corresponding authors: vishnukumar@iisc.ac.in , saurabhc@iisc.ac.in

**ABSTRACT:**

Piezoelectric micro-electro-mechanical systems have significant market potential owing to their superior capabilities of transduction to those of standard capacitive and piezoresistive devices. However, piezoelectric films are often lossy, which reduces the quality factor of devices and affects their performance. It is thus important to examine all sources of energy losses in such devices and accurately determine them based on experimental data. Currently used methods to quantify energy loss from different sources and the properties of materials based on experimental data are set-up for piezoelectric devices, in which energy storage and loss primarily occur in the same piezoelectric material. Moreover, such methods rely on resonance-antiresonance measurements, and thus are unsuitable for thin-film-piezoelectric-on-substrate (TPoS) micro/nano devices that have i) a significant portion of energy stored in the substrate/device layer, ii) a low signal-to-noise ratio owing to either lossy piezoelectric films or low motional impedance, or iii) a larger feedthrough capacitance in addition to the internal capacitance of the piezoelectric film. In this paper, we propose a method that overcomes these challenges based on synchronized optical and electrical measurements. We develop a comprehensive physics-based model to extract all the relevant parameters for the device, including the coefficient of piezoelectric coupling, internal and feedthrough capacitance, loss tangents (dielectric, piezoelectric, and mechanical), and the contributions of different sources to the quality factor of the device. We showcase the proposed method by using a PZT-based TPoS MEMS cantilever. It can be universally applied to all piezoelectric materials and arbitrary stacks of the device layer.

**INTRODUCTION**

Piezoelectric materials such as quartz, lead zirconate titanate (PZT), lead magnesium niobate-lead titanate (PMN-PT), and aluminum nitride have a wide range of applications, including for RF timing reference, sensors based on surface acoustic waves, bulk acoustic waves, and energy harvesting [1]–[6]. Quality factor, a measure of energy loss in a resonator, is an important quantity that determines performance of MEMS/NEMS devices [7]. Thus, to design and engineer the Quality Factor (Q), it is important to study the energy loss mechanisms prevalent in the materials constituting the MEMS/NEMS device. Energy loss mechanisms in piezoelectric materials have been extensively studied since the 1950s [8]–[12].

The three main mechanisms of energy loss that have been identified in piezoelectric materials are dielectric loss, mechanical loss, and piezoelectric loss [13]. Dielectric loss occurs due to the finite resistance of the piezoelectric film leading to irreversible resistive losses. As a result of the finite resistance, the electric displacement lags behind the applied electric field by a phase lag, $\delta$, the tangent of which (tan$\delta$) is known as dielectric loss [14]. Mechanical losses include all material losses, such as thermoelastic dissipation, the Akhieser effect, and viscoelastic losses [15]. This loss occurs due to a phase lag, $\phi$, between the stress and the strain, and its corresponding loss tangent is tan$\phi$ [8], [16]. Piezoelectric losses occur due to a phase lag between the electric field and the mechanical strain, and vice versa. This phase lag occurs owing to the finite relaxation time required for the adjustment of the dipoles after the application of the electric field. It is denoted by $\theta$, and the associated loss tangent is tan$\theta$ [8]. Holland et.al. introduced loss tangents as complex coefficients for various properties of the piezoelectric materials to encapsulate the underlying physics of the individual mechanism of the energy loss in a single parameter [17], [18]. A summary of these mechanisms of energy loss in TPoS micro/nano-electro-mechanical system (M/NEMS) devices is shown in Fig. 1.

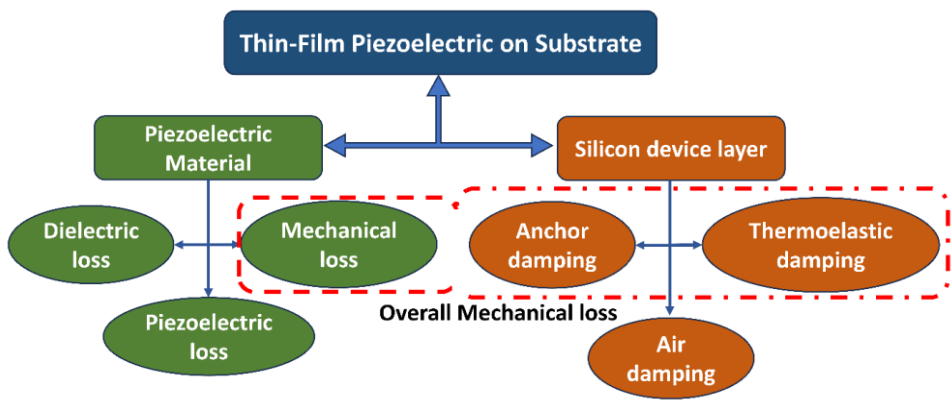

Fig. 1. Energy loss mechanisms in TPoS MEMS/NEMS devices.

From an engineering standpoint, it is important to quantify various material properties and energy loss mechanisms in piezoelectric devices to enable geometry and device optimization. In bulk materials, it is possible to study the loss tangents individually using direct measurements viz capacitance-conductance measurements, indentation, force hammer method, and laser interferometry [12], [16], [19]–[21]. However, for thin film-based devices, while direct C-V measurements can be readily performed, the other measurements are harder to carry out owing to complexities introduced due to the stiffness of the substrate and the electrostatic contribution to displacement in addition to that resulting from piezoelectric actuation [22]–[24].

Thus, several indirect methods for extracting various loss parameters using impedance measurements at resonance and anti-resonance in conjunction with simulations, (e.g., the IEEE Std. 176 [25]), have been explored. One such modelling approach was proposed by Smits [26]. As per this approach, an initial guess of real and imaginary parts of the compliance coefficient is made. By fitting the impedance data for two frequencies using a model for a bar of piezoelectric material in conjunction with the initial guess, the relative dielectric constant and piezoelectric coefficients are calculated. These calculated values are further used to determine the compliance coefficient at a different frequency leading to a new initial guess. The iterations are continued till the compliance value converges [26], [27]. The drawbacks of such approaches are:

1) If the guess is poorly chosen, especially in the case of lossy piezoelectric materials, convergence takes a lot more iterations [28].

2) If the output signal is low and the antiresonance falls below the noise floor of the circuit, this method is rendered ineffective.

3) A feedthrough signal from the source to the sensing electrode affects the 3dB bandwidth of the resonance and antiresonance peaks.

4) Finally, the major drawback of this method is that this formulation is applicable only for devices with a stack primarily consisting of the piezoelectric material, and the majority of energy stored and lost is in the same material. However, in the TPoS platform, often the device layer on which piezoelectric material is deposited is thick and stores a significant amount of the energy while the energy loss occurs in the entire stack.

To work around this fundamental issue, Afzaal et al. suggested an approach that requires fabricating several separate devices with different incremental stacks to estimate the energy loss contributions due to the individual films [29]. This is particularly onerous since the fabrication of devices with several different material stacks is necessary. Furthermore, this method still does not account for the pervasive feedthrough in TPoS devices.

To address all these challenges in estimating quality factor contributions of individual films of lossy TPoS devices while accounting for the feedthrough, we present a new measurement and modelling technique that requires the fabrication of devices with a single stack. The measurement technique involves simultaneous optical and electrical measurement carried out at only two frequencies, one at resonance and another at a sufficiently low frequency compared to the resonance frequency, using a Laser Doppler Vibrometer (LDV) and lock-in amplifier (LIA). We present a physics-based analytical model for Piezo-SOI resonant devices that includes the effect of all sources of energy loss as well as feedthrough capacitance. With the aid of this model, we demonstrate that using the aforementioned measurement on a single device, it is possible to extract all relevant parameters such as piezoelectric coupling coefficient ($e_{31}$), internal as well as feedthrough capacitance ($C_p$ & $C_f$), and all loss tangents ($\tan\theta$, $\tan\delta$, & $\tan\phi$). We demonstrate our model by showing measurement results for a cantilever beam fabricated using TPoS technology that encompasses actuation and sensing simultaneously and extracts the parameters as stated above. These measurements were carried out using two different configurations that exhibit differing feedthrough capacitance, and our model correctly distinguishes between the two. Using these extracted parameters, the entire resonant frequency response for displacement and voltage is derived and shown to match very well with the displacement and voltage spectral response of the device. Furthermore, we show that an independent measurement of dielectric loss tangent of the piezoelectric material matches exactly with the extracted parameters, thus, further affirming the veracity of our model and measurement technique.

**NUMERICAL METHOD**

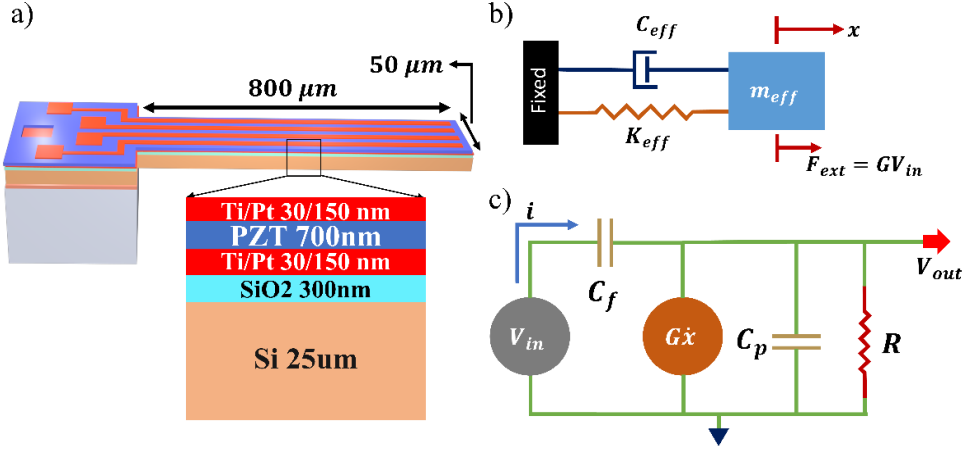

Fig. 2. a. Schematic of a TPoS MEMS cantilever, b. A mechanical equivalent system of the cantilever for force balance, where external force is generated through piezoelectric transduction c. An electrical equivalent system for charge balance including rate of charge developed from strain in the cantilever.

We fabricated a piezo-SOI cantilever by using the process outlined by Tiwari et al. [30]. A schematic of the design of this device is shown in Fig. 2a, which shows that a PZT film with a thickness of 700 nm was used for piezoelectric transduction. When a voltage was applied, a force was generated by the piezoelectric effect that acted on the system. We can model it as a simple spring mass damper system, as shown in Fig. 2b [31].

The force-balance equation is written as:

$$M_{eff}\ddot{x} + \eta_{eff}\dot{x} + K_{eff}x - GV_{in} = 0 \qquad (1)$$

where, $M_{eff}$, $\eta_{eff}$, and $K_{eff}$ are the effective mass, damping coefficient and stiffness that are calculated using the geometry, boundary conditions and material composition of the stack that makes up the cantilever system, $G$ is the electromechanical coupling coefficient, and $V_{in}$ is the actuation voltage. Here, the damping factor $\eta_{eff}$ accounts for the air damping, dielectric, and piezoelectric damping. All other losses are independently accounted for as discussed subsequently.

Similarly, due to the direct piezoelectric effect, a charge develops across the sensing electrodes on the same cantilever as a result of strain developed by the bending of the cantilever. The equivalent electrical model is depicted in Fig. 2c. The charge-balance equation for the circuit is written as:

$$(C_p + C_f)\dot{V}_{out} + \frac{V_{out}}{R} + G\dot{x} - C_f\dot{V}_{in} = 0 \qquad (2)$$

Here, $C_p$ is the capacitance of the piezoelectric material, $C_f$ is the feedthrough capacitance induced due to the field lines between actuation and sensing electrodes and $R$ is the load resistance at the output end of the measurement unit.

The electromechanical coupling coefficient is a function of the electromechanical constant ($e_{31,f}$) and specific geometry of the device.

$$G = b_{eff} \cdot e_{31,f}^* \cdot t \cdot \vartheta'(x)|_{x=l} \tag{3}$$

$b_{eff}$, $t$, and $\vartheta(x)$ are the geometrical parameters, i.e., width of the electrode, distance from neutral axis of the cantilever, and shape function of the cantilever in the particular mode of actuation (refer Appendix A), and $e_{31}$, f is the effective electromechanical constant.

The piezoelectric material's stiffness, electromechanical constant, and capacitance can be written in terms of phase delays. Piezoelectric material stiffness is related to the compliance (S) of the material and compliance can be written in the form of elastic phase delay as [13]:

$$S^* = S_0 \cdot (1 - jtan\phi) \tag{4}$$

where $\phi$ is defined as phase angle between the stress and the strain.

Electromechanical constant ($e_{31}^* = \frac{d_{31}^*}{S^*}$) can be written in form of electromechanical phase delay as:

$$d_{31}^* = d_{31}^0 \cdot (1 - jtan\theta) \tag{5a}$$

$$e_{31}^* = e_{31}^0 \cdot (1 - j(tan\theta - tan\phi)) \tag{5b}$$

where, $\theta$ is defined as the phase delay between the applied electric and the strain.

The dielectric constant can be written as:

$$\varepsilon^* = \varepsilon^0 \cdot (1 - jtan\delta) \tag{6}$$

where, $\delta$ is defined as the phase delay between the electric field and the electric displacement. Using the above mentioned basic complex forms, corresponding dependence of the loss tangents on the stiffness, electromechanical constant, and capacitance can be written as:

$$K_{eff} = K_0 \cdot (1 + jtan\phi) \tag{7a}$$

$$e_{31,f}^* = e_{31,f} \cdot (1 - j(tan\theta - tan\phi)) \tag{7b}$$

$$C_p = C_p^0 \cdot (1 - jtan\delta) \tag{7c}$$

where, $K_0$ is the cantilever stiffness calculated by COMSOL simulation.

The coupled Equations (1) and (2) can be solved using Laplace transforms and the solution to the mechanical response and the output voltage response to an applied input voltage is shown in Equations (8) and (9) (details of the calculations are included in Appendix A).

$$\frac{X}{V_{in}} = \frac{G_o \cdot [1 - j(\tan\theta - \tan\phi)]}{\left(-M_{eff}\omega^2 + \eta_{eff} \cdot j\omega + K_o \cdot (1 + j\tan\phi)\right)} \quad (8)$$

$$\frac{V_{out}}{V_{in}} = \frac{\left[C_f - G_o^2 \cdot (1 - 2j(\tan\theta - \tan\phi)) \cdot \left\{\frac{1}{-M_{eff}\omega^2 + K_o + j(K_o\tan\phi + \omega\eta_{eff})}\right\}\right] \cdot j\omega}{\frac{1}{R} + \omega C_p^0 \tan\delta + j\omega(C_p^0 + C_f)} \quad (9)$$

These solutions can be simplified in three different regimes, viz., $\omega \ll \omega_{res}$, $\omega = \omega_{res}$, and $\omega \gg \omega_{res}$. They are summarized in Table I.

**Table I: mechanical and electrical response as a function of various parameters in different actuation frequency regimes.**

|  | $\omega \ll \omega_{res}$ | $\omega = \omega_{res}$ | $\omega \gg \omega_{res}$ |
|---|---|---|---|
| $\dfrac{X}{V_{in}}$ | $\dfrac{G_o}{K_o}(1 - jtan\theta)$ | $\dfrac{G_o}{\eta_{eff}\omega + K_o\tan\phi}(\tan\phi - \tan\theta)\left[1 - \dfrac{j}{\tan\theta - \tan\phi}\right]$ | $\dfrac{G_o}{M_{eff}\omega^2}\left[1 - j\left(\tan\theta - \tan\phi - \dfrac{K_o\tan\phi}{M_{eff}\omega^2} - \dfrac{\eta_{eff}}{M_{eff}\omega}\right)\right]$ |
| $\dfrac{V_{out}}{V_{in}}$ | $\dfrac{C_f}{C_p^0}[1 + jtan\delta]$ | $\dfrac{1}{C_p^0}\left[C_f + \dfrac{G_o^2}{K_o\tan\phi + \omega\eta_{eff}} \times (-j - 2\tan\theta) + jC_f\tan\delta\right]$ | $\dfrac{1}{C_p^0}\left[C_f + \dfrac{G_o^2}{-M_{eff}\omega^2}\left\{1 - 2j(\tan\theta - \tan\phi) + j\dfrac{(K_o\tan\phi + \omega\eta_{eff})}{M_{eff}\omega^2}\right\} + jC_f\tan\delta\right]$ |

We note that in the simplified forms of the system response in the regimes $\omega \ll \omega_{res}$ and $\omega = \omega_{res}$, the highly coupled dependencies on loss tangents and various material properties decompose into much more tractable versions that can be used to extract individual material property parameters. Using these simplified forms, we developed an algorithm for estimating $e_{31,f}$, $C_p$, $C_f$, $\eta_{eff}$, $\tan\delta$, $\tan\theta$, and $\tan\phi$ (shown in Fig. 3).

Here, we note that it is generally observed that the responses of $\dfrac{X}{V_{in}}$ and $\dfrac{V_{out}}{V_{in}}$ have very noisy phase in the regime $\omega \ll \omega_{res}$ making it impossible to accurately detect the phase. On the other hand, the amplitude can be readily estimated with fair accuracy. Thus, in the regime $\omega \ll \omega_{res}$, the algorithm utilizes only the amplitude of the response for the requisite estimation of parameters while ignoring the phase component.

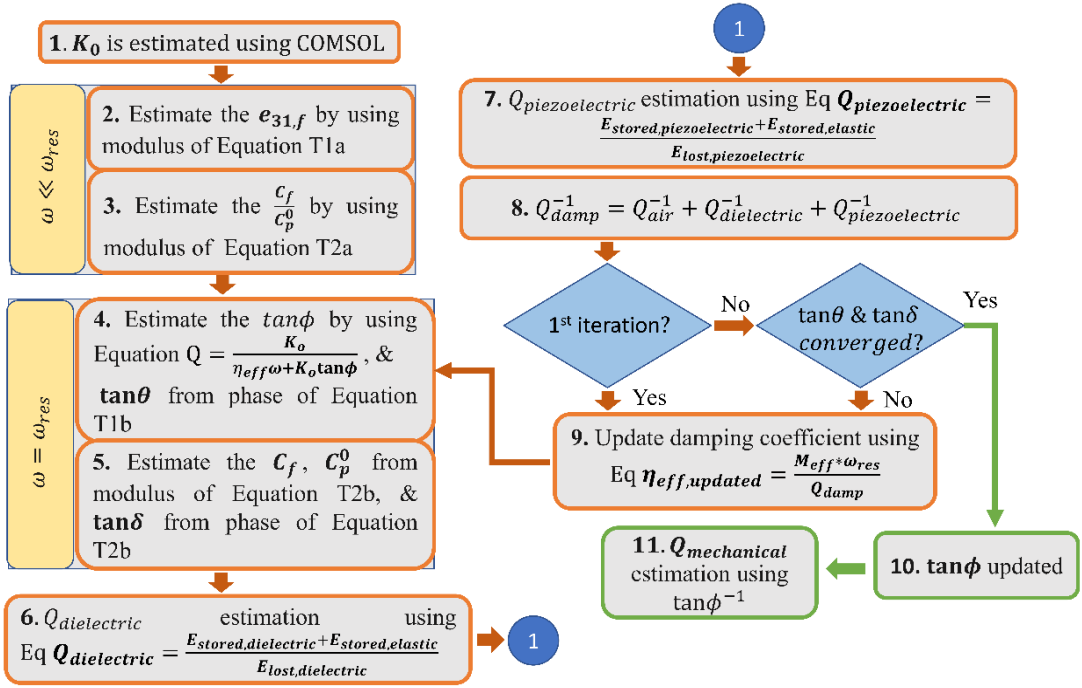

Fig. 3. Algorithm to estimate the piezoelectric parameters $e_{31,f}$, $C_p$, $C_f$, $\eta_{eff}$, $\tan\delta$, $\tan\theta$, and $\tan\phi$, and the quality factors $Q_{dielectrics}$, $Q_{piezoelectric}$, and $Q_{mechanical}$.

As per step 1 of the algorithm shown in Fig. 3, we begin by estimating $K_0$ through COMSOL by applying a point load at the tip of the cantilever. In step 2, we estimate $G_0$ and calculate the effective piezoelectric coupling coefficient ($e_{31,f}$). The amplitude of voltage response in the regime $\omega \ll \omega_{res}$ can be used to calculate the ratio $\frac{C_f}{C_p^0}$ as described in step 3.

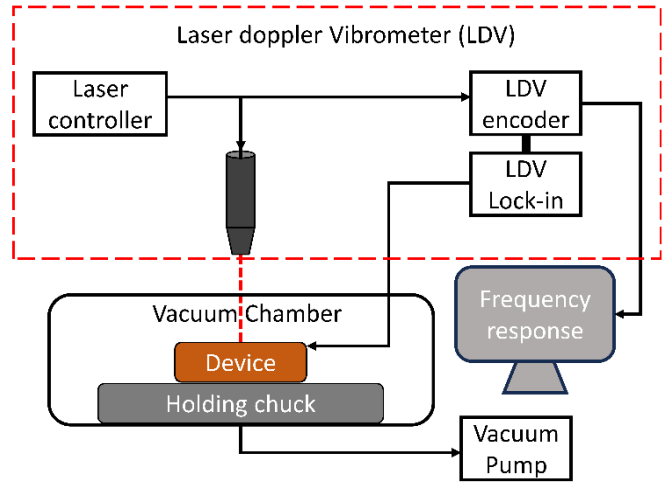

Fig. 4. LDV based estimation of air damping using vacuum chamber setup.

$Q_{air}$ is estimated using Equation (10) and used to calculate $\eta_{eff}$ using Equation (14b) and the experimental setup for measurement of $Q_{air,ldv}$ and $Q_{vacuum,ldv}$ is shown in Fig. 4.

$$Q_{air}^{-1} = Q_{ldv\ in\ air}^{-1} - Q_{ldv\ in\ vacuum}^{-1} \qquad (10)$$

The overall quality factor, $Q$, is estimated using the LDV measurement at resonant frequency ($\omega \ll \omega_{res}$) and off resonant low frequency ($\omega = \omega_{res}$) and applying Equation (26). $\tan\phi$ can be calculated by step 4 of the algorithm. Note that this initially calculated value of $\tan\phi$ is an overestimate since it also accounts for piezoelectric and dielectric losses and needs to be corrected via iterations. We then use the phase component of the displacement response to the input voltage at the resonant frequency to calculate $\tan\theta$. We then calculate $C_f$, $C_p^0$ and $\tan\delta$ as per step 5, and $Q_{dielectric}$ per step 6. Following this, we calculate $Q_{piezoelectric}$ and the updated $Q_{damping}$ following steps 7 and 8. We then update $\eta_{eff}$ as prescribed by step 9 (refer Appendix B), and repeat the estimation loop from step 4 onward. This iteration is stopped when $\tan\delta$ and $\tan\theta$ converge to a value within an acceptable tolerance. Thus, the final estimate for $\tan\phi$ contains components of pure mechanical loss while the other losses are separately accounted for.

**EXPERIMENTAL SETUP**

The schematic of a typical measurement setup is shown in Fig. 5a, and an optical image of the device is shown in Fig. 5b. The cantilever device was driven by using an input signal on the outermost electrode (D). The sinusoidal output signal generated by using the signal generator of the LDV was passed through a splitter to provide a reference signal for the lock-in (Zurich MFLI), while the other split signal was connected to the device to drive the cantilever. The output signal (S) from outermost electrode was connected to the lock-in input to obtain an amplitude and phase relative to the reference signal. This output signal was plotted by using Zurich LabOne Software. Optical measurements of the mechanical response were simultaneously carried out from the cantilever system. The forced signal input at a specified frequency was then converted into an FFT signal in LDV software to obtain the mechanical response at the given frequency. These optical and electrical measurements were carried out at different frequencies to obtain the responses in the frequency spectrum. Here, we note that though our algorithm requires measurements at only two frequencies, a complete experimental response was measured to compare against the output response predicted by the model. In addition, LDV measurements were carried out in a vacuum to isolate the component of energy loss due to air damping, discussed in the numerical method section. These measurements were carried out by using a termination resistance of 1 MΩ.

The measurement configuration described above used a single input and generated a single output (SISO), with the two central electrodes grounded to reduce feedthrough. A double-input double-output (DIDO) configuration as shown in Fig. 6 could also be implemented. In this case, a differential drive signal was applied to one of the inner electrodes and an outer electrode, while a differential sensing signal was obtained from the remaining inner electrode and outer electrode. Unlike the case of a balanced, fully differential device, the feedthrough was higher in the DIDO configuration because the differential drive and sense were not balanced. This was subsequently demonstrated by using the experimental results and is reflected in our proposed model.

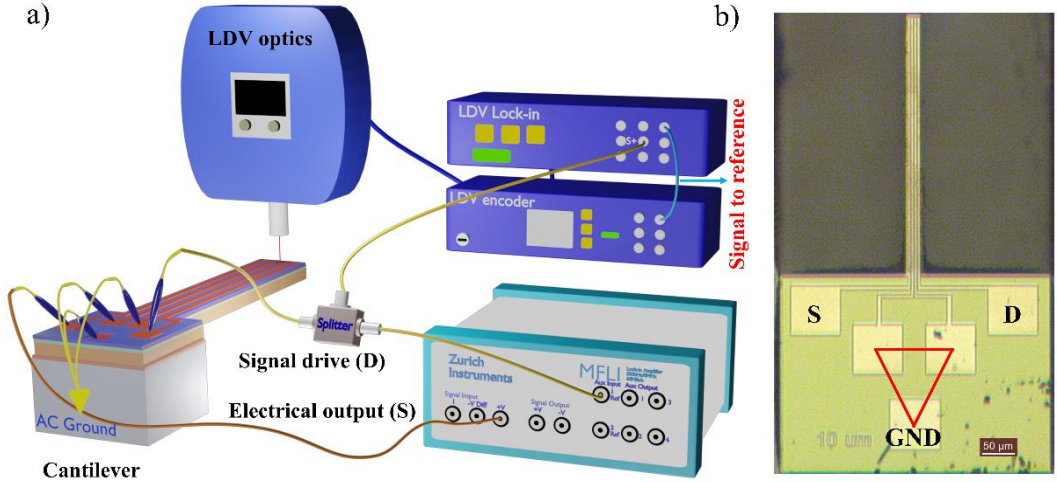

Fig. 5. a. Single Input Single Output (SISO) synchronized Opto-Electro-Mechanical measurement schematic. Two outer electrodes are used for drive and sense, while the other two inner electrodes are grounded to reduce crosstalk between probed electrodes, b. optical image of the TPoS device.

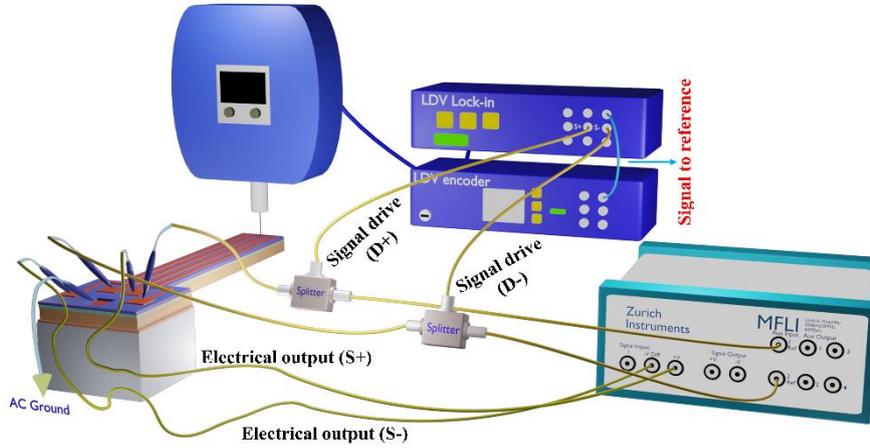

Fig. 6. Double Input Double Output (DIDO) synchronized Opto-Electro-Mechanical measurement schematic. One outer electrode at positive (D+) and one inner electrode at negative (D-) are used for drive, while the other outer and inner electrodes are used for sense.

**RESULTS**

A. Single Input Single Output (SISO)

The line schematic for the SISO configuration of transduction is shown in Fig. 7a. The mechanical and electrical responses at a drive voltage of 0.2 V for this configuration are shown in Fig. 7b and 7c respectively. The algorithm described in Fig. 3 was applied for data points collected in the two regimes $\omega \ll \omega_{res}$ and $\omega = \omega_{res}$ and the parameters extracted using these two points were used to generate a complete model for the system. Fig. 7b and 7c also show the comparison of the experimentally obtained data vs the model. In the subsequent subsection, the procedure for extraction of the parameters is explained in further detail for the SISO case. The estimated requisite parameters and the overall quality factor, as well as the estimated contributions of the individual mechanisms of energy loss to the quality factor under the SISO setup are listed in Tables II and III, respectively.

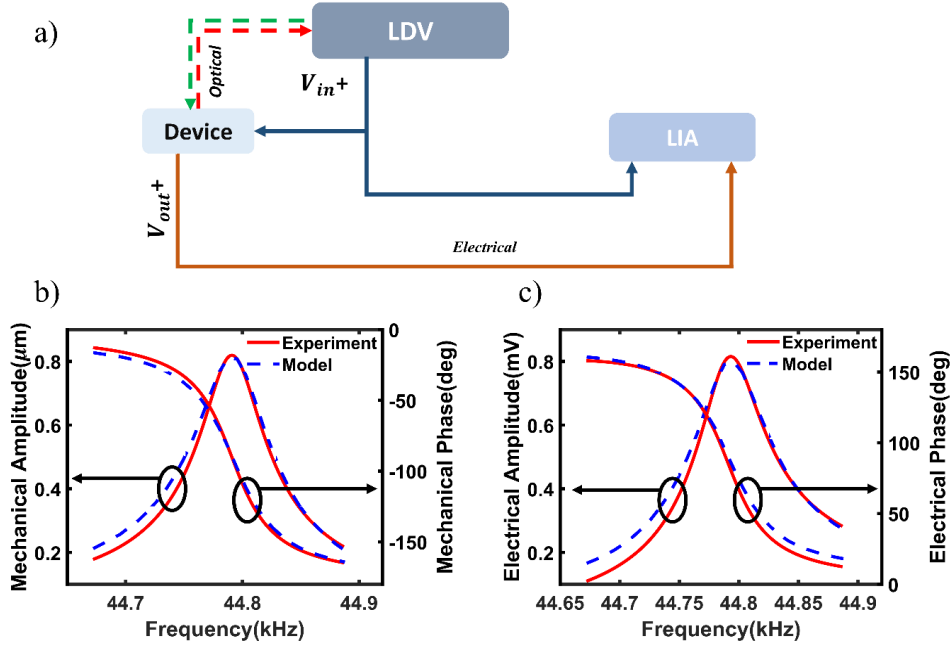

Fig. 7. Single Input and Single Output Configuration a. Line schematic of the configuration, b. Mechanical response using optical detection of LDV and estimated response obtained using our model (standard deviation for amplitude and phase measurements are $5.6 \cdot 10^{-3}$ µm and $0.39°$ respectively), c. Electrical response measured using the LIA and fitted with estimated response obtained via our model (standard deviation for amplitude and phase measurements are $1.22 \cdot 10^{-3}$ mV and $0.072°$ respectively).

$Q_{TED}$ was calculated by using the COMSOL thermoelasticity module. $Q_{other}$, which represents anchor loss and the other mechanisms of mechanical energy loss, was estimated by using Equation (11).

$$Q_{other}^{-1} = Q_{mechanical}^{-1} - Q_{TED}^{-1} = tan\phi - Q_{TED}^{-1} \qquad (11)$$

The calculated frequency response for the entire frequency range in which the measurements were performed is shown in Figures 7b and 7c. As can be seen, even though the model only uses inputs from two frequencies, it predicts the entire frequency response extremely well in correspondence to the measured data.

**Table II. Estimated parameters.**

| $e_{31,f}$ (C/m²) | $C_p^0$ (F) | $C_f$ (F) | $tan\phi$ | | $tan\theta$ | $tan\delta$ |
|---|---|---|---|---|---|---|
| | | | $tan\phi_{pzt}$ | $tan\phi_{other}$ | | |
| $-1.31$ | $3.9 \times 10^{-10}$ | $5.8 \times 10^{-14}$ | $1.31 \times 10^{-5}$ | $3.17 \times 10^{-4}$ | $9.4 \times 10^{-3}$ | $2.62 \times 10^{-2}$ |

**Table III. Quality factors.**

| $Q_{experiment}$ | $Q_{air}$ | $Q_{dielectric}$ | $Q_{piezoelectric}$ | $Q_{mechanical}$ | |
|---|---|---|---|---|---|
| | | | | $Q_{TED}$ | $Q_{others}$ |
| 845 | 1478 | 5818.2 | $2.57 \times 10^5$ | 17854 | 3647.9 |

B. Double Input Double Output (DIDO)

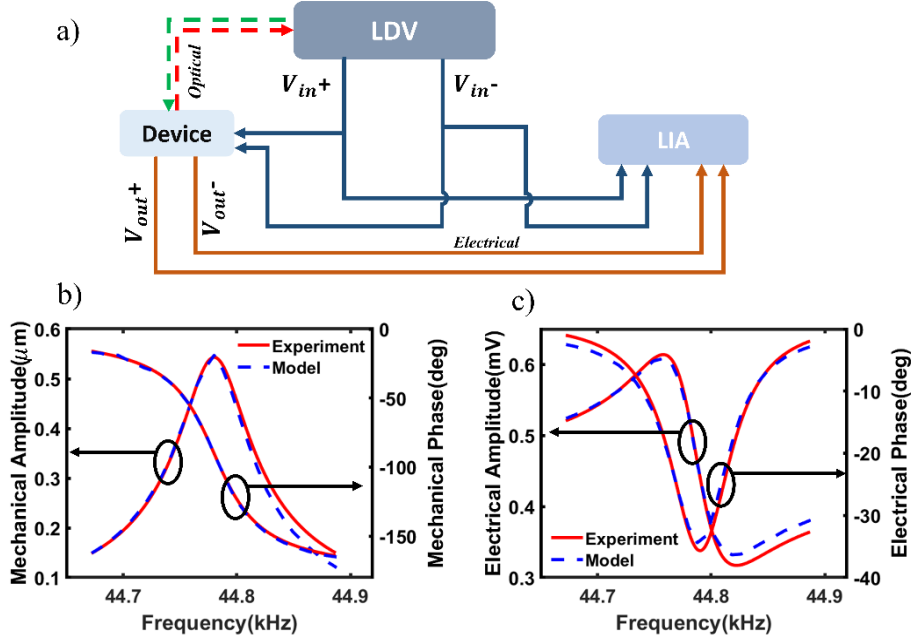

Fig. 8. Double Input and Double Output Configuration a. Line schematic of the configuration, b. Mechanical response using optical detection of LDV and estimated response obtained using our model (standard deviation for amplitude and phase measurements are $2.78 \cdot 10^{-2}$ µm and $1.18^o$ respectively), c. Electrical response measured using the LIA and fitted with estimated response obtained via our model (standard deviation for amplitude and phase measurements are $9.6 \cdot 10^{-4}$ mV and $0.09^o$ respectively).

In the DIDO configuration, the voltage input scheme is shown in Fig. 8a. A differential input voltage was applied to one of the inner and the outer electrodes on the other side. Fig. 8b and 8c show the measurements and the model's prediction for this configuration. The model parameters were estimated in the same way as the SISO case using the prescribed algorithm in Fig. 3. In the differential configuration, the outer electrode and the inner electrode exert equal and opposite force, but their relative efficacy in moving the cantilever is not equal. Thus, though $e_{31}$ remains unchanged for SISO and DIDO configurations, G values are different. One can account for this change using an effective width, $b_{eff}$, instead of the width of a single strip of the electrode (b). The estimated parameters for this configuration are shown in Table IV, and the quality factor estimates are shown in Table V.

## Table IV: Estimated parameters

| $b_{eff}$ (μm) | $e_{31,f}$ (C/m²) | $C_p^0$ (F) | $C_f$ (F) | $tan\phi$ | | $tan\theta$ | $tan\delta$ |
|---|---|---|---|---|---|---|---|
| | | | | $tan\phi_{pzt}$ | $tan\phi_{other}$ | | |
| 4.6 | $-1.31$ | $4.96 \times 10^{-10}$ | $1.09 \times 10^{-12}$ | $1.31 \times 10^{-5}$ | $2.47 \times 10^{-4}$ | $8.93 \times 10^{-3}$ | $2.43 \times 10^{-2}$ |

## Table V: Quality factors

| $Q_{experiment}$ | $Q_{air}$ | $Q_{dielectric}$ | $Q_{piezoelectric}$ | $Q_{mechanical}$ | |
|---|---|---|---|---|---|
| | | | | $Q_{TED}$ | $Q_{others}$ |
| 740 | 1478 | 2435.6 | $2.3 \times 10^5$ | 17854 | 4907 |

## DISCUSSION

A. Electrical output with respect to velocity

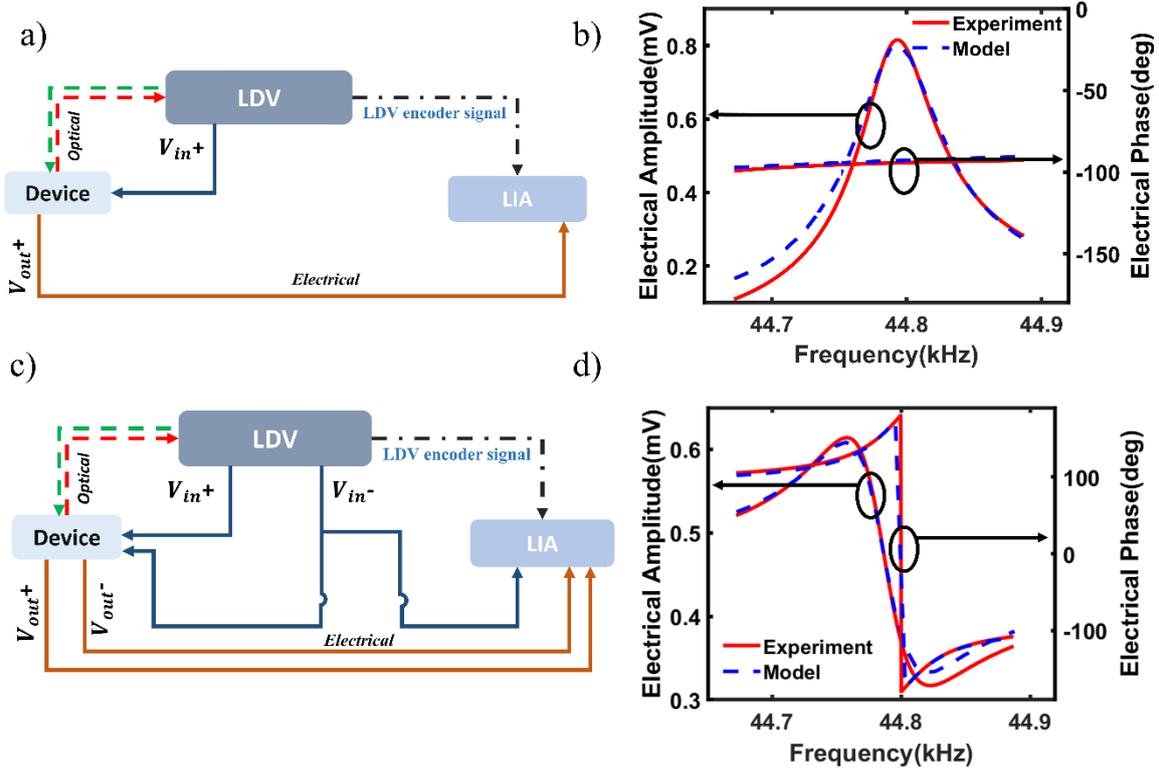

Fig. 9. Voltage to velocity response of SISO configuration and DIDO configuration a. the line diagram of SISO showing the connection from the output signal of LDV to the auxiliary port of LIA, b. Measured electrical response for the SISO configuration from the LIA and the estimated response calculated via our model. c. The line diagram of DIDO shows the connection from the output signal of LDV to the auxiliary port of LIA, d. Measured electrical response for the DIDO configuration and the estimated response obtained from our model.

In order to test the model, we conducted another measurement wherein the output of the LDV encoder, which carries a signal proportional to the velocity of the device, was used as a reference to measure the relative phase and amplitude of the electrical output according to the Equation (19). The configuration of measurement in the SISO system is shown in Fig. 9a and the measured data along with the model's prediction is shown in Fig. 9b. It can be seen that the model fits the measured data very well. A similar measurement was performed for the DIDO system and the details of this measurement setup are shown in Fig. 9c. The measured data and the model's excellent prediction are shown in Fig. 9d.

B. Experimental verification of tan$\delta$

The dielectric loss tangent is an essential parameter to determine the energy loss in the dielectric material. Our estimated tan$\delta$ value for the device is 0.0262. We have characterized this material in past work [32] by using capacitance-conductance measurement through a semiconductor device analyzer. This direct measurement yielded a value of 0.026. This shows that our algorithm for parameter estimation is highly accurate and robust, given that the electrical and mechanical responses are intricately dependent on various parameters coupled with one another and we, nonetheless, obtained an accurate estimate of the dielectric loss.

C. Feedthrough capacitance estimation

As mentioned above, the feedthrough capacitance is often ignored by the standard techniques for estimating film properties. Our model highlights the effects of feedthrough capacitance on the device performance, and this was also reflected in the measurements. The feedthrough capacitance was higher in the DIDO system because the outer sensing electrode was in the direct vicinity of the inner drive electrode, with insufficient cancelling differential signal induced by the distant outer drive electrode. On the contrary, the inner electrodes were grounded in the SISO system to effectively sever any field connection between the two outer electrodes. Thus, the estimated feedthrough capacitance for the SISO system was $5.8 \cdot 10^{-14}$ $F$, while that for the DIDO system is $1.09 \cdot 10^{-12}$ F.

D. $e_{31,f}$ estimation

Several researchers have developed methods to determine the effective transverse piezoelectric coefficients, either based on pure measurement [33] or through a combination of measurements and models [34], [35]. These measurements have been conducted either for standalone PZT film or for TPoS devices [33], [34], [36], [37]. Our algorithm provides an accurate estimate of the $e_{31,f}$ coefficient. The estimated value of the effective transverse piezoelectric coefficient for the SISO method of actuation was $-1.31 \frac{C}{m^2}$. An independent measurement of the $d_{33}$ coefficient was carried out by using a Berlincourt meter for the same piezoelectric film and was found to be 52.6±4.3 $\frac{pm}{V}$. Thus, with the assumption that $d_{31} = -0.5 d_{33}$ [38] and elastic modulus as specified in Table VI, the independently measured estimate of $e_{31,f}$ was $-1.29 \frac{C}{m^2}$. Our estimate of $e_{31}$ was remarkably accurate with less than 1% error. We also observe that the value of $e_{31}$ of our film was lower than typically reported values in the range of -1.78 $\frac{C}{m^2}$ to -5 $\frac{C}{m^2}$ [39]–[42]. This shows that the quality of the film that we used in our experiments was worse than those considered for measurements in the past works.

E. Quality factor estimation

Minimizing energy dissipation in a device can enhance the quality factor and improve its performance. In our case, we needed to develop a new approach to estimate the components of energy loss because our PZT material was poor in quality. Our approach also enables the determination of the components of quality factors corresponding to various mechanisms, such as $Q_{piezoelectric}$, $Q_{dielectric}$ and $Q_{mechanical}$, in the TPoS system without the need to fabricate several devices with stacks of different material. An interesting result of this study is that increasing energy stored in a thick silicon device layer substantially improves the quality factor of the system.

**CONCLUSION**

In conclusion, we have developed a method for estimating the piezoelectric parameters of TPoS devices. These parameters are challenging to measure by using conventional methods, such as the IEEE standard 176, without using a multitude of devices with stacks of different films. We use a synchronized opto-electro-mechanical measurement of a single TPoS device in conjunction with a model that accounts for the feedthrough capacitance. We also reported a systematic algorithm for estimating all the relevant piezoelectric parameters and the contributions of all individual mechanisms of energy loss to the quality factor, including $e_{31}$, $C_p$, $C_f$, $\tan\theta$, $\tan\phi$, $\tan\delta$, $Q_{dielectric}$, $Q_{piezoelectric}$, and $Q_{mechanical}$. We verified the accuracy of our technique of measurement and our algorithm by using a TPoS cantilever device operated in the SISO and DIDO configurations. Our model requires only two measurements of the response of the resonator, viz., at $\omega = \omega_{res}$ and $\omega \ll \omega_{res}$. We showed that the proposed model can accurately measure the complete mechanical and electrical responses with respect to the input signal as well as the output voltage response with respect to the velocity of the resonator beam at all frequencies. Furthermore, we verified that the value of $\tan\delta$ obtained from the model matched the value determined by using a separate C-G measurement within 1%. Moreover, we showed that our method accounts for the feedthrough capacitance, which has a significant impact on the electrical performance of the resonator, as is evident from the results of the system for DIDO measurements. We were able to calculate the coupling coefficient for the film without using the resonanceantiresonance measurement for our poor-quality film, which further demonstrates the robustness of our technique. Our results showed that the estimated $e_{31}$ was within 1% of its measured value, where this further attests to the accuracy of our model. Finally, we estimated the components of $Q$ corresponding to all the major sources of energy loss, without the need to fabricate devices with permutations of stacked materials, which is an expensive and time-consuming endeavor.

**ACKNOWLEDGMENT**

We wish to acknowledge the support of Centre for Nanoscience and Engineering, IISc, and the National Nano Fabrication Centre (NNFC) and Micro-Nano Characterization Facility (MNCF). Vishnu Kumar gratefully acknowledge the MHRD, Govt. of India for providing us the necessary funding and fellowship to pursue research work. Saurabh Chandorkar acknowledges Indian Space Research Organization, Government of India (GoI) under Grant

DS 2B13012(2)/41/2018- Sec.2, by the Ministry of Electronics and Information Technology, GoI under 25(2)/2020 ESDA DT.28.05.2020 and Ministry of Human Resource and Development, GoI Grant SR/MHRD 18 0017. Saurabh Chandorkar also acknowledges DRDO JATP: DFTM/02/3125/M/12/MNSST-03.

**APPENDIX A**: Model of mechanical and electrical transduction in thin-film piezoelectric-on-substrate (TPoS) devices, and its solution

In order to estimate various individual parameters relevant for the TPoS devices, it is necessary to develop a comprehensive model for the mechanical and electrical transduction. From force balance, using Equation (1):

$$M_{eff}\ddot{x} + \eta_{eff}\dot{x} + K_{eff}x - GV_{in} = 0 \tag{12}$$

From rate of charge balance, using Equation (2):

$$(C_p + C_f)\dot{V}_{out} + \frac{V_{out}}{R} + G\dot{x} - C_f\dot{V}_{in} = 0 \tag{13}$$

where,

$$G = b_{eff} \cdot e^*_{31,f} \cdot t \cdot \vartheta'(x)|_{x=l} \tag{14a}$$

$$\eta_{eff} = \frac{M_{eff} \cdot \omega}{Q_{air}} \tag{14b}$$

$$\vartheta(x) = \{\cos(\alpha x) - \cosh(\alpha x)\} - \frac{\cos(\alpha l) + \cosh(\alpha l)}{\sin(\alpha l) + \sinh(\alpha l)}\{\sin(\alpha x) - \sinh(\alpha x)\} \tag{15}$$

$$\alpha l = 1.875 \; for \; 1st \; mode \; for \; cantilever$$

As mentioned in the Numerical Method section, various loss parameters can be incorporated in the model by using a complex formulation of various material properties as per

$$K_{eff} = K_0 \cdot (1 + j\tan\phi) \tag{16a}$$

$$e^*_{31,f} = e_{31,f} \cdot \left(1 - j(\tan\theta - \tan\phi)\right) \tag{16b}$$

$$C_p = C_p^0 \cdot (1 - j\tan\delta) \tag{16c}$$

$$G = G_o \cdot [1 - j(\tan\theta - \tan\phi)] \tag{16d}$$

Here, we have assumed a single $\phi$ for the overall elastic modulus of the material stack. This phase delay not only includes thermoelastic dissipation but also incorporates clamping loss and other intrinsic loss mechanisms of loss. The Laplace transform of force balance Equation (12) can be used to separate the displacement with respect to input voltage as

$$\frac{X}{V_{in}} = \frac{G}{(M_{eff}s^2 + \eta_{eff}s + K_{eff})} \tag{17}$$

and output voltage Equation (13) as below:

$$\frac{V_{out}}{V_{in}} = \frac{C_f s}{\frac{1}{R} + s(C_p + C_f)} - \frac{Gs}{\frac{1}{R} + s(C_p + C_f)} \cdot \frac{X}{V_{in}} \tag{18}$$

The output voltage with respect to velocity can be written as:

$$\frac{V_{out}}{\dot{X}} = \frac{\frac{V_{out}}{V_{in}}}{\left(s \cdot \frac{X}{V_{in}}\right)} \tag{19}$$

where, $s = j\omega$ for the harmonic response.

By expanding Equation (17) through the substitution of terms from Eqns. (16a, b, c, and d), we get:

$$\frac{X}{V_{in}} = \frac{G_o \cdot [1 - j(\tan\theta - \tan\phi)]}{\left(-M_{eff}\omega^2 + \eta_{eff}j\omega + K_o \cdot (1 + j\tan\phi)\right)} \tag{20}$$

We now consider three separate regimes of frequency of excitation viz. $\omega \ll \omega_{res}$, $\omega = \omega_{res}$ and $\omega \gg \omega_{res}$.

**Case 1: $\omega \ll \omega_{res}$**

For $\omega \ll \omega_{res}$, Equation (20) simplifies to

$$\frac{X}{V_{in}} = \frac{G_o}{K_o} \cdot \left[1 - j\left(\tan\theta + \frac{\omega\eta_{eff}}{K_o}\right)\right] \tag{21}$$

Assuming $\frac{\omega\eta_{eff}}{K_o} \ll \tan\theta$

$$\frac{X}{V_{in}} = \frac{G_o}{K_o} \cdot (1 - j\tan\theta) \tag{22}$$

We note here that the assumption $\frac{\omega\eta_{eff}}{K_o} \ll \tan\theta$ is not required for the algorithm we have presented since we don't use the phase measured off resonance as it is usually noisy.

***Case*2: $\omega = \omega_{res}$**

For $\omega = \omega_{res}$, Equation (20) simplifies to

$$\frac{X}{V_{in}} = \frac{G_o}{\eta_{eff}\omega + K_o\tan\phi}[-(\tan\theta - \tan\phi) - j] \tag{23}$$

Writing the equation in a manner such that the phase can be extracted easily, we get,

$$\frac{X}{V_{in}} = \frac{G_o}{\eta_{eff}\omega + K_o\tan\phi}(\tan\phi - \tan\theta)\left[1 - \frac{j}{\tan\phi - \tan\theta}\right] \tag{24}$$

By noting that $\tan\phi - \tan\theta \ll 1$, the amplitude of Equation (24) simplifies to

$$\left|\frac{X}{V_{in}}\right| = \frac{G_o}{\eta_{eff}\omega_{res} + K_o\tan\phi} \qquad (25)$$

We can calculate the Quality factor, $Q$ using standard expression for a linear second order system,

$$Q = \frac{\left|\frac{X}{V_{in}}\right|_{\omega=\omega_{res}}}{\left|\frac{X}{V_{in}}\right|_{\omega\ll\omega_{res}}} \qquad (26)$$

$$Q = \frac{\frac{G_o}{\eta_{eff}\omega + K_o\tan\phi}}{\frac{G_o}{K_o}} = \frac{K_o}{\eta_{eff}\omega + K_o\tan\phi} \qquad (27)$$

### Case 3: $\omega \gg \omega_{res}$

We also consider the case $\omega \gg \omega_{res}$ for completeness since it is not required for the proposed algorithm. Equation (20) simplifies to

$$\frac{X}{V_{in}} = \frac{G_o \cdot [1 - j(\tan\theta - \tan\phi)]}{-M_{eff}\omega^2\left[1 - j\left(\frac{K_o\tan\phi + \omega\eta_{eff}}{M_{eff}\omega^2}\right)\right]} \qquad (28)$$

Upon simplifying the complex component, we get,

$$\frac{X}{V_{in}} = \frac{G_o}{-M_{eff}\omega^2}\left[1 - j\left(\tan\theta - \tan\phi - \frac{K_o\tan\phi}{M_{eff}\omega^2} - \frac{\eta_{eff}}{M_{eff}\omega}\right)\right] \qquad (29)$$

From Equation (18), substituting Equations (20), and (16), we get,

$$\frac{V_{out}}{V_{in}} = \frac{\left[C_f - G_o^2 \cdot (1 - 2j(\tan\theta - \tan\phi)) \cdot \left\{\frac{1}{-M_{eff}\omega^2 + K_o + j(K_o\tan\phi + \omega\eta_{eff})}\right\}\right] \cdot j\omega}{\frac{1}{R} + \omega C_p^0\tan\delta + j\omega(C_p^0 + C_f)} \qquad (30)$$

As before, we consider three separate regimes of frequency of excitation, viz. $\omega \ll \omega_{res}$, $\omega = \omega_{res}$ and $\omega \gg \omega_{res}$.

### Case 1: at $\omega \ll \omega_{res}$

For $\omega \ll \omega_{res}$, Equation (30) reduces to

$$\frac{V_{out}}{V_{in}} = \frac{(C_f + \frac{G_o^2}{K_o})}{(C_p^0 + C_f)}\left[1 - j\left\{\left(\frac{\frac{G_o^2}{K_o}}{C_f + \frac{G_o^2}{K_o}}\right) \cdot \left(-2\tan\theta + 2\tan\phi - \tan\phi - \frac{\omega\eta_{eff}}{K_o}\right) + \left(\frac{\frac{1}{\omega R} + C_p^0\tan\delta}{C_p^0 + C_f}\right)\right\}\right] \qquad (31a)$$

$$\frac{V_{out}}{V_{in}} = \frac{C_f}{C_p^0 + C_f}\left[1 - j\left\{\frac{G_o^2}{K_oC_f} \cdot \left(-2\tan\theta + \tan\phi - \frac{\omega\eta_{eff}}{K_o}\right) + \left(\frac{\frac{1}{\omega R} + C_p^0\tan\delta}{C_p^0 + C_f}\right)\right\}\right] \qquad (31b)$$

Assuming, $C_f \gg \frac{G_o^2}{K_o}$, which is generally true for TPoS devices with thick device layers,

We note that this assumption is not really necessary for the algorithm, but it does simplify the calculations to some extent. Furthermore assuming, $\frac{1}{R} \ll \omega C_p^0 \tan\delta$, we get

$$\frac{V_{out}}{V_{in}} = \frac{C_f}{C_p^0 + C_f}\left[1 + j\left(\frac{C_p^0 \tan\delta}{C_p^0 + C_f}\right)\right] \quad (32)$$

This result is not used for the algorithm suggested since the measured phase component for off-resonance excitation is noisy and, is hence ignored.

In general, it is usually true that $C_f \ll C_p^0$ and incorporating this assumption yields,

$$\frac{V_{out}}{V_{in}} = \frac{C_f}{C_p^0} \cdot [1 + j\tan\delta] \quad (33)$$

### *Case*2: $\omega = \omega_{res}$

For excitation at resonance frequency, i.e. $\omega = \omega_{res}$, Equation (30) simplifies to

$$\frac{V_{out}}{V_{in}} = \frac{C_f - \frac{G_o^2}{K_o \tan\phi + \omega\eta_{eff}} \cdot (-j - 2(\tan\theta - \tan\phi))}{(C_p^0 + C_f) \cdot \left[1 - j\left(\frac{\frac{1}{\omega R} + C_p^0 \tan\delta}{C_p^0 + C_f}\right)\right]} \quad (34)$$

From displacement response, $\tan\theta > \tan\phi$ and $\frac{1}{\omega R} \ll C_p^0 \cdot \tan\delta$

$$\frac{V_{out}}{V_{in}} = \frac{C_f - \frac{G_o^2}{K_o \tan\phi + \omega\eta_{eff}} \cdot (-j - 2\tan\theta)}{(C_p^0 + C_f) \cdot \left[1 - j\left(\frac{C_p^0 \tan\delta}{C_p^0 + C_f}\right)\right]} \quad (35)$$

which, under the aforementioned assumption of $C_f \ll C_p^0$, reduces to

$$\frac{V_{out}}{V_{in}} = \frac{C_f - \frac{G_o^2}{K_o \tan\phi + \omega\eta_{eff}} \cdot (-j - 2\tan\theta)}{(C_p^0 + C_f) \cdot [1 - j\tan\delta]} \quad (36)$$

By calculating the amplitude of Equation (36), we arrive at

$$(C_p^0 + C_f)^2 \cdot \left|\frac{V_{out}}{V_{in}}\right|^2_{\omega=\omega_{res}} = C_f^2 - \frac{4G_o^2 C_f \tan\theta}{(K_o \tan\phi + \omega\eta_{eff})} + \frac{G_o^4}{(K_o \tan\phi + \omega\eta_{eff})^2} \quad (37)$$

$C_f$ and $C_p^0$ are estimated by Equation (37) combining with amplitude of Equation (33). Phase of the Equation (36) provide $\tan\delta$, Furthermore,

$$\tan\left(\angle\left(\frac{V_{out}}{V_{in}}\right)\right) = \frac{C_f \tan\delta - \frac{G_o^2}{K_o \tan\phi + \omega\eta_{eff}}}{C_f - \frac{2G_o^2 \tan\theta}{(K_o \tan\phi + \omega\eta_{eff})} + \frac{G_o^2 \tan\delta}{(K_o \tan\phi + \omega\eta_{eff})}} \qquad (38)$$

Assuming, $\tan\left(\angle\left(\frac{V_{out}}{V_{in}}\right)\right) = p$, and $\frac{G_o^2}{K_o \tan\phi + \omega\eta_{eff}} = q$, Equation (38) can be rearranged to calculate $\tan\delta$

$$\tan\delta = \frac{p \times C_f - 2\tan\theta \times p \times q + q}{C_f - p \times q} \qquad (39)$$

**Case 3**: $\omega \gg \omega_{res}$

We also consider the case $\omega \gg \omega_{res}$ for completeness since it is not required for the proposed algorithm. Equation (30) simplifies to

$$\frac{V_{out}}{V_{in}} = \frac{\left[C_f + \frac{G_o^2}{-M_{eff}\omega^2}(1 - 2j(\tan\theta - \tan\phi)) \times \left\{\frac{1}{1 - j * \frac{(K_o \tan\phi + \omega * \eta_{eff})}{M_{eff}\omega^2}}\right\}\right]}{(C_p^0 + C_f) \times \left[1 - j\left(\frac{\frac{1}{\omega R} + C_p^0 \tan\delta}{C_p^0 + C_f}\right)\right]} \qquad (40)$$

Using the same assumptions mentioned in the case of $\omega = \omega_{res}$, we get,

$$\frac{V_{out}}{V_{in}} = \frac{1}{C_p^0}\left[C_f + \frac{G_o^2}{-M_{eff}\omega^2}\left\{1 - 2j(\tan\theta - \tan\phi) + j\frac{(K_o \tan\phi + \omega\eta_{eff})}{M_{eff}\omega^2}\right\}\right.$$
$$\left. + jC_f \tan\delta\right] \qquad (41)$$

**APPENDIX B**: Updating $\eta_{eff}$

As noted in the algorithm, the initial assumption for $\eta_{eff}$ is calculated using $Q_{air}$ alone as $Q_{piezoelectric}$ and $Q_{dielectric}$ have not been calculated or measured. Thus,

$$\eta_{eff-initial} = \frac{M_{eff}\omega_{res}}{Q_{air}} \qquad (42)$$

In subsequent iterations, when $Q_{piezoelectric}$ and $Q_{dielectric}$ are calculated as per the algorithm, $\eta_{eff}$ needs to be updated as

$$\eta_{eff,updated} = \frac{M_{eff}\omega_{res}}{Q_{damp}} \qquad (43)$$

where,

$$Q_{damp}^{-1} = Q_{air}^{-1} + Q_{dielectric}^{-1} + Q_{piezoelectric}^{-1} \qquad (44)$$

Here, $Q_{dielectric}$ is estimated by using the dielectric loss and elastic stored energy using direct vibration amplitude [32] as:

$$Q_{dielectric} = \frac{Energy_{stored,dielectric} + Energy_{stored,elastic}}{Energy_{lost,dielectric}} \quad (45)$$

where,

$$Energy_{stored,elastic} = \frac{1}{2} K_0 \left(\frac{A_{mechanical}}{Q_{air}} \times Q_{medium}\right)^2 \quad (46)$$

$$Energy_{stored,dielectric} = \frac{1}{2} C_{p,electrode}^0 V_{in}^2 \quad (47)$$

$$Energy_{lost,dielectric} = Energy_{stored,dielectric} \times \tan\delta \quad (48)$$

Here, $C_{p,electrode}^0$ is the capacitance at the electrode, $A_{mechanical} = \left|\frac{X}{V_{in}}\right|_{\omega=\omega_{res}}$, $Q_{medium} = Q_{air}$, as LDV was performed in air.

Furthermore, $Q_{piezoelectric}$ was estimated by taking loss of electromechanical energy into account as per procedure outlined in [8].

$$Q_{piezoelectric} = \frac{Energy_{stored,piezoelectric} + Energy_{stored,elastic}}{Energy_{lost,piezoelectric}} \quad (49)$$

$$Energy_{stored,piezoelectric} = \frac{1}{2} \frac{e_{31,f}^2}{E_{piezo}} E_0^2 \times Vol. \quad (50)$$

where, $E_0^2$ is electric field supplied to the device, and $E_{piezo}$ is Young's modulus of piezoelectric film.

$$Energy_{lost,piezoelectric} = Energy_{stored,piezoelectric} \times (2\tan\theta - \tan\phi) \quad (51)$$

**Table VI:** Properties [43] used for simulations and estimations.

| Parameters | Details | Value |
|---|---|---|
| $E_{Si}$ | Elastic modulus of Si | 170 [GPa] |
| $E_{SiO2}$ | Elastic modulus of SiO$_2$ | 70 [GPa] |
| $E_{Pt}$ | Elastic modulus of Pt | 168 [GPa] |
| $E_{PZT}$ | Elastic modulus of PZT | 49 [GPa] |
| $\rho_{Si}$ | Density of Si | 2329 [kg/m$^3$] |
| $\rho_{SiO2}$ | Density of SiO$_2$ | 2200 [kg/m$^3$] |
| $\rho_{Pt}$ | Density of Pt | 21450 [kg/m$^3$] |
| $\rho_{PZT}$ | Density of PZT | 7500 [kg/m$^3$] |
| $\nu_{Si}$ | Poisson's ratio of Si | 0.28 |
| $\nu_{SiO2}$ | Poisson's ratio of SiO$_2$ | 0.17 |
| $\nu_{Pt}$ | Poisson's ratio of Pt | 0.38 |
| $\nu_{PZT}$ | Poisson's ratio of PZT | 0.31 |
| $\alpha_{Si}$ | Thermal expansion coefficient of Si | $2.6 \times 10^{-6}$ [1/K] |
| $\alpha_{SiO2}$ | Thermal expansion coefficient of $SiO_2$ | $5 \times 10^{-7}$ [1/K] |
| $\alpha_{Pt}$ | Thermal expansion coefficient of Pt | $8.8 \times 10^{-6}$ [1/K] |
| $\alpha_{PZT}$ | Thermal expansion coefficient of PZT | $3.5 \times 10^{-6}$ [1/K] |
| $k_{Si}$ | Thermal conductivity of Si | 130 [W/m/K] |
| $k_{SiO2}$ | Thermal conductivity of $SiO_2$ | 1.4 [W/m/K] |
| $k_{Pt}$ | Thermal conductivity of Pt | 71.6 [W/m/K] |
| $k_{PZT}$ | Thermal conductivity of PZT | 1.2 [W/m/K] |
| $Cp_{Si}$ | Heat capacity at constant pressure of Si | 700 [J/Kg×K] |
| $Cp_{SiO2}$ | Heat capacity at constant pressure of $SiO_2$ | 730 [J/Kg×K] |
| $Cp_{Pt}$ | Heat capacity at constant pressure of Pt | 133 [J/Kg×K] |
| $Cp_{PZT}$ | Heat capacity at constant pressure of PZT | 420 [J/Kg×K] |
| Lc | Length of the cantilever | 800 [μm] |
| Wc | Width of the cantilever | 50 [μm] |
| $t_{cantilever}$ | Thickness of the cantilever | 26.2 [μm] |
| b | Width of the electrode | 6 [μm] |